\documentclass{webofc}
\usepackage[varg]{txfonts}   
\woctitle{19th International Seminar on High Energy Physics QUARKS-2016}
\begin{document}

\begin{flushright}
INR-TH-2016-026
\end{flushright}

\title{Systematic treatment of non-linear effects in \\
Baryon Acoustic Oscillations }
%
%

\author{\firstname{Mikhail M.} \lastname{Ivanov}\inst{1,2,3}\fnsep\thanks{\email{mikhail.ivanov@cern.ch}} 
}

\institute{FSB/IPHYS/LPPC, \'Ecole Polytechnique F\'ed\'erale de Lausanne,  CH-1015, Lausanne, Switzerland
\and
Institute for Nuclear Research of the
Russian Academy of Sciences,
60th October Anniversary Prospect, 7a, 117312
Moscow, Russia
\and
Department of Particle Physics and Cosmology, Physics Faculty, Moscow State University, Vorobjevy Gory, 119991, Moscow, Russia
          }

\abstract{
In this contribution we will discuss the non-linear effects in the baryon acoustic oscillations and present a systematic and controllable
way to account for them within {\it time-sliced perturbation theory}.  
}
\maketitle
\section{Introduction}
\label{intro}

Baryon acoustic oscillations (BAO)
are widely used to establish the distance-redshift relation
enabling to constrain the expansion history and the composition of the Universe \cite{Eisenstein:1998tu,Eisenstein:2005su}. 
BAO measurements aim at providing (sub-) percent precision in the near future and thus
it is imperative to provide the best possible theoretical control over the BAO 
in order to fully exploit the potential of future surveys.

\begin{figure}
\begin{center}
  \includegraphics[width=0.49\textwidth]{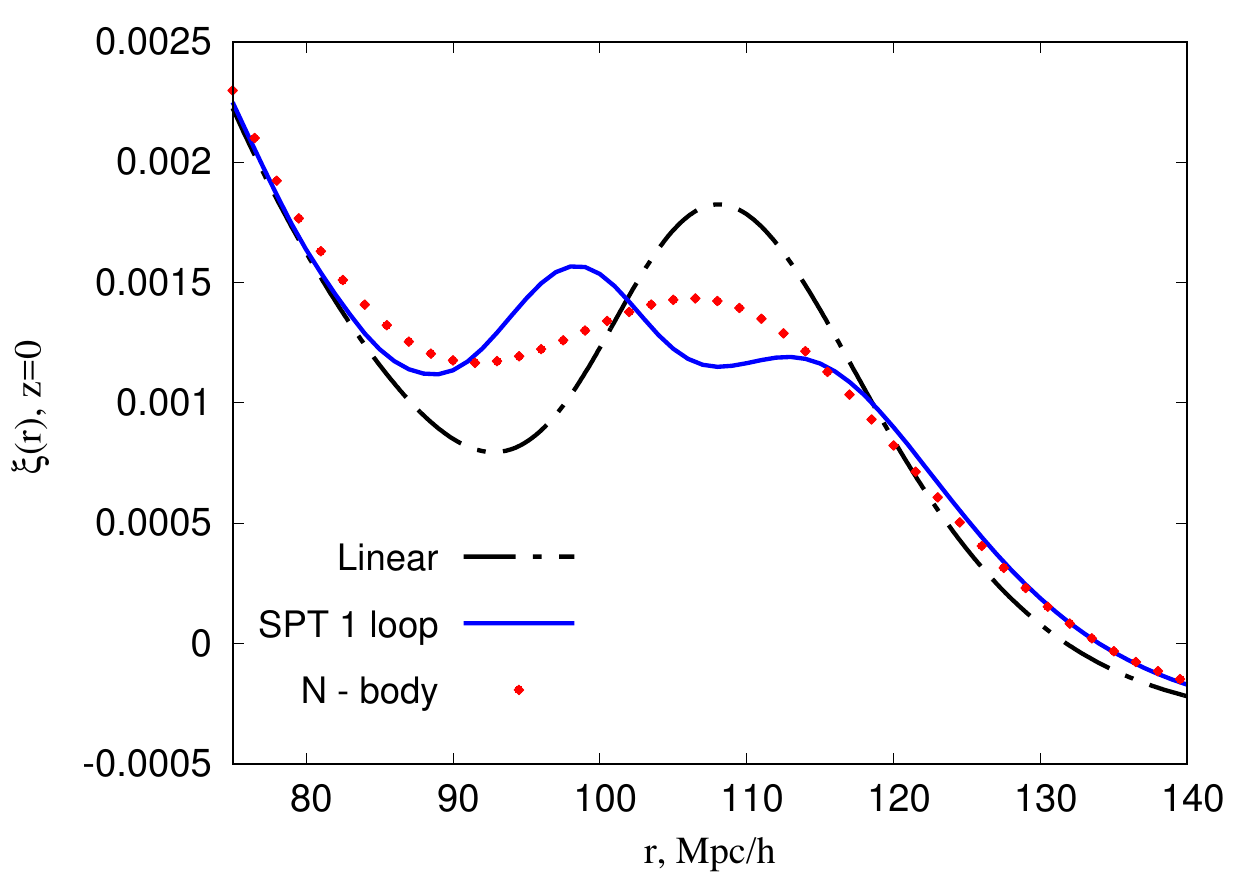}
    \includegraphics[width=0.49\textwidth]{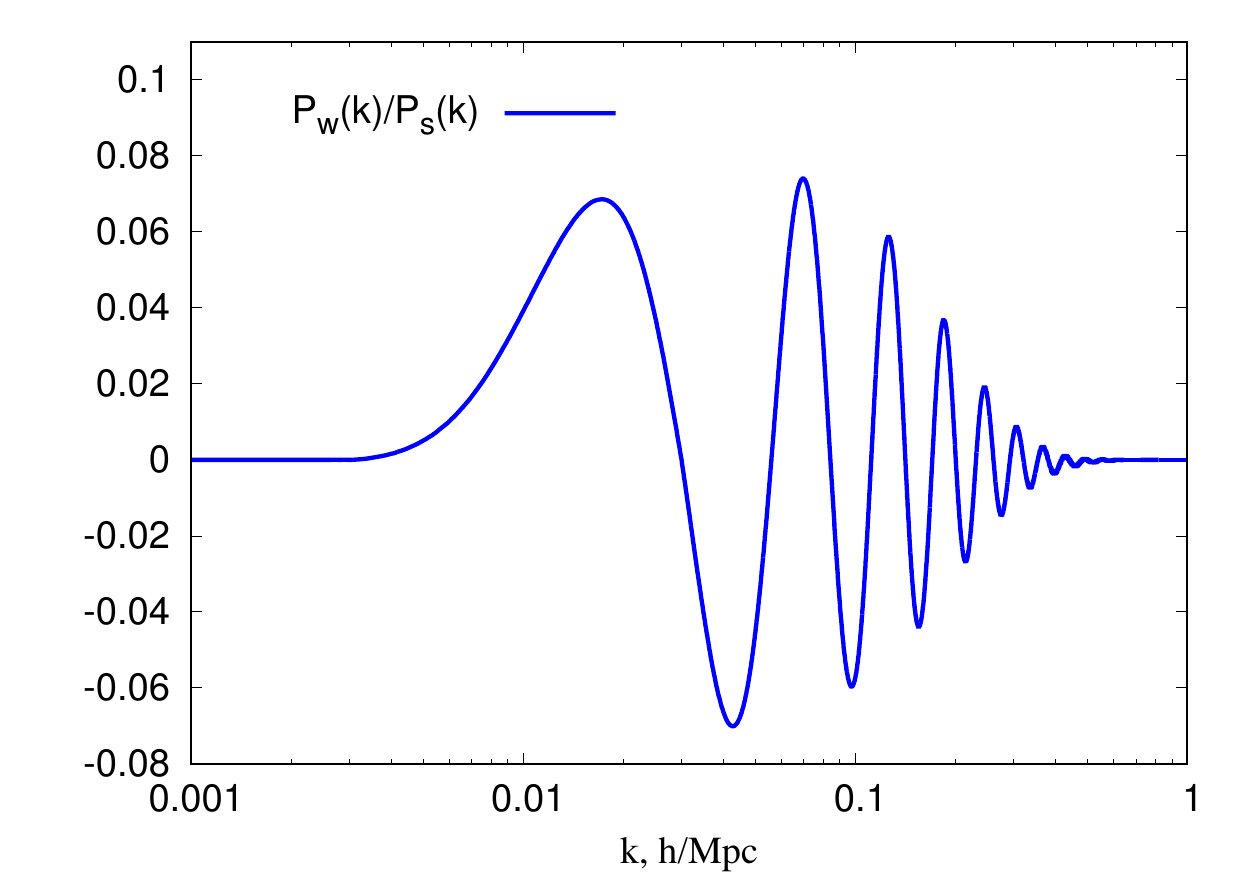}
\end{center}
\caption{\label{fig1} 
{\it Left panel}: Two-point matter correlation functions in position space at redshift zero, $\xi(r)\equiv \langle \delta(\textbf{r})\delta(0)\rangle$:
N-body data from the Horizon Run 3 simulation \cite{Kim:2011ab} (red points), the prediction of linear theory (black dot-dashed line), and the 1-loop SPT result (blue solid line).
{\it Right panel}:
Ratio of oscillatory (wiggly) part $P_w$ of the linear power spectrum
to the smooth part $P_s$.
The $\Lambda$CDM
cosmological parameters have been chosen as in \cite{Kim:2011ab}.
}
\end{figure}

The BAO are most prominent in the 2-point correlation function of matter density in 
position space where they form a peak at $r_{BAO}\sim $110 Mpc$/h$ in comoving coordinates.
It has been observed long ago that the shape of the BAO peak retrieved from N-body simulations 
differs significantly from the prediction of linear theory (see Fig.~\ref{fig1}, left panel).
Even though $r_{BAO}$ is significantly larger than the characteristic scale of non-linear clustering $2\pi k_{NL}^{-1}\sim 20$ Mpc$/h$,
the leading correction computed in Eulerian standard perturbation theory (SPT) failed to capture the behavior 
seen in N-body data (see Fig.~\ref{fig1}, left panel). 
This disagreement is caused by large-scale bulk flows whose interaction with short modes is amplified 
if the distribution of matter has a feature.
At leading order the effect of bulk motions is to worsen the correlation between galaxies at separations of order $r_{BAO}$, which results in the suppression of the BAO peak.

Many approaches have been put forward in order to account for bulk flows. 
From the numerical side one can mention BAO {\it reconstruction} \cite{Eisenstein:2006nk}, 
which undoes bulk motions directly in the data and yields a sharper
BAO peak with the signal-to-noise ratio improved by a factor of 2. 
From the theoretical side among the most successful approaches 
we would like to mention renormalized perturbation theory \cite{CrSc1,Crocce:2007dt}, Lagrangian perturbation theory \cite{Matsubara:2007wj}
and IR - resummed effective field theory of large scale structures \cite{Senatore:2014via,Baldauf:2015xfa}.
In this short contribution we will present a new way to systematically describe the non-linear evolution of BAO in the 
framework of time-sliced perturbation theory (TSPT) \cite{Blas:2015qsi,Blas:2016sfa}.

\section{Interaction with bulk flows: a simple physical picture}
\label{sec:phys_pict}

In this section we present a simple intuitive picture behind the non-linear evolution of the BAO 
(see Refs \cite{Crocce:2007dt,Matsubara:2007wj,Senatore:2014via,Baldauf:2015xfa,Blas:2015qsi,Blas:2016sfa,Tassev:2013rta,Sherwin:2012nh} for more details).
We start with the density contrast field in Fourier space, $\delta(\textbf{k},\eta)\equiv (\rho(\eta,\textbf{k})-\bar \rho(\eta))/\bar \rho(\eta)$,
where $\eta$ is conformal time, $\rho(\eta,\textbf{k})$ is the local density and $\bar \rho(\eta)$ is the average matter density in the Universe.
$\delta(\eta,\textbf{k})$ is a random stochastic field and one is typically interested in its 2-point correlation function 
(power spectrum) defined as $P(\eta;k)\delta^{(3)}_{\text{Dirac}}(\textbf{k}+\textbf{k}')\equiv \langle \delta(\eta,\textbf{k})\delta(\eta,\textbf{k}')\rangle$.
The linear power spectrum in our Universe can be decomposed into the smooth and wiggly parts (see Fig.~\ref{fig1}, right panel), 
\begin{equation}
P_{\text{lin}}(\eta;k)=P_s(\eta;k)+P_w(\eta;k)\,,
\end{equation}
and these two parts receive non-linear corrections from long-wavelength perturbations (IR modes) in very different ways.

Let us first focus on the situation in which the linear power spectrum has only the smooth part, i.e. is featureless.
Consider two galaxies on top of a large scale bulk flow.
If the separation between these galaxies is shorter than the long wavelength associated to the flow, 
then at zeroth order its effect is a uniform acceleration, $\textbf{a}=-\nabla \Phi_L(x)\simeq$ const ($\Phi_L(x)$ is a gravitational potential 
induced by a long-wavelength perturbation).
Then, according to the equivalence principle, at a given time the effect of this acceleration can be totally removed by a proper coordinate transformation,
so that it cannot affect equal-time physical observables such as the power spectrum.
At next-to-leading order one can take into account inhomogeneity in the acceleration, in other words, the fact that the flow is not uniform but rather "converging" or "diverging". In that case two galaxies are likely to be found at 
separations shorter ("converging" flows) or larger ("diverging" flows)
than in linear theory.
However, since the smooth power spectrum in the real Universe is close to scale-invariant, 
the effect of the motions due to long modes should be small.
In SPT the latter fact is manifest if one considers the 1-loop correction to the power spectrum in the IR limit 
($q\ll k$, where $q$ is the loop momentum),
\begin{equation}
\label{eq:SPT1loopIR}
P_{\text{1-loop, SPT}}(k)=P_{\text{lin}}(k)+\left(\frac{569}{735}P_{\text{lin}}(k)-\frac{47}{105}kP'_{\text{lin}}(k)+\frac{1}{10}k^2P''_{\text{lin}}(k)\right)\sigma^2_S\,,
\end{equation}
where $\sigma^2_S = \int_0^{q_{\text{max}}\ll k}d^3q P_{\text{lin}}(q)\ll 1$ is the large scale variance of the density field. 
From Eq.\eqref{eq:SPT1loopIR} it is clear that the interaction between the mode with momentum $k$ and the IR modes is suppressed provided that 
the derivatives of the linear power spectrum are small, 
$k^nP^{(n)}_{\text{lin}}\sim P_{\text{lin}}$.
Thus, expanding over the coupling to the IR modes ($\sim \sigma^2_S$) yields a converging perturbative expansion in this case.
As the reader might have already guessed, this is not the case for a power spectrum with a feature.

Now consider what happens in the presence of the wiggly component. 
In that case the effect of non-uniform accelerations is not small anymore.
Indeed, the presence of a feature at
$r_{BAO}\equiv k_{osc}^{-1}$ means, at the linear level, that two galaxies 
have bigger probability to be found at this separation.
At the non-linear level bulk flows make these galaxies move to shorter ("converging" flows) or larger ("diverging" flows) separations,
which results in the degradation of the correlation
between them at $r_{BAO}$ once these motions are averaged over long modes.
Since matter is collapsing, "converging" flows are 
more common and two galaxies should be typically found at separations systematically smaller than $r_{BAO}$.
This effect is known as the {\it physical shift} of the BAO peak \cite{Sherwin:2012nh,Smith:2007gi} and this is a next-to-leading order effect as compared to the broadening due to bulk flows.
Assuming $P_w(k)=~f_s(k)\sin(k/k_{osc})$ ($f_s(k)$ is a monotonic function) 
the SPT 1 loop result yields (here we assume $q\leq q_{max}\ll k_{osc}$),
\begin{subequations}
 \begin{align}
P_{w,\text{1-loop, SPT}}(k)&=P_{w}(k)+O(1)\times \sigma^2_S P_w(k)-\frac{k}{k_{osc}}\left(\frac{47}{105}-\frac{kf'_s}{5f_s}\right)\sigma^2_S f_s\cos(k/k_{osc})-\frac{k^2}{k_{osc}^2}\frac{\sigma^2_S}{10}P_w\,,\\
\label{eq:SPT1loopIRw}
&= f_s(k)\left(1-\frac{k^2}{k_{osc}^2}\frac{\sigma^2_S}{10}\right) \sin
\left(\frac{k}{k_{osc}}\left\{ 1-\left[\frac{47}{105}-\frac{kf'_s}{5f_s}\right]\sigma^2_S \right\}\right) +O(1)\times \sigma^2_S P_w(k)
+O( \sigma^4_S )\,.
\end{align}
\end{subequations}
From \eqref{eq:SPT1loopIRw} one clearly sees the mentioned effects 
on acoustic oscillations: 
the suppression of their amplitude and the shift 
in their phase w.r.t linear theory. The effective coupling constant to the IR modes $\sim \sigma^2_S(k^2/k^2_{osc})$ is parametrically enhanced compared to the featureless case and in fact is of order unity for realistic cosmology. Thus, the perturbative expansion as \eqref{eq:SPT1loopIRw} does not make sense unless one 
can resumm relevant contributions to all orders in perturbation theory. 
We will show now how 
this procedure, called {\it IR-resummation} can be systematically performed in TSPT.

\section{Overview of Time-Sliced Perturbation Theory}

Time - sliced perturbation theory is a novel approach to large-scale structure (LSS) formation aimed at studying equal-time correlators in perturbation theory
\cite{Blas:2015qsi}. 
It is instructive to focus first on the SPT-like approach, in which one studies the following partition function\footnote{Typically in LSS one studies the density and velocity divergence fields, however, only one of them is statistically independent and can be an argument of the partition function in Eq.\eqref{eq:Zspt}.
In Ref. \cite{Blas:2016sfa} we chose this to be the velocity field.
In this paper, for presentation reasons, we switched to the density contrast~$\delta$.
One can show that the resummation procedure in the case of this variable will be completely similar to the one performed in~\cite{Blas:2016sfa}.
 },
\begin{equation}
\label{eq:Zspt}
Z[J]=\mathcal{N}^{-1}\int \mathcal{D}\delta_0 \; \exp \left\{-\frac{1}{2}\int d^3k \frac{\delta_0(\textbf{k})\delta_0(-\textbf{k})}{P_0(k)}+\int d^3k \;\delta_\eta(\textbf{k})J(-\textbf{k})\right\}\,,
\end{equation}
where the ensemble average is taken over the initial field that has a Gaussian distribution. 
The time-dependence of the field $\delta(\eta,\textbf{k})$ is governed by the system of the continuity, Euler and Poisson equations 
for a perfect pressurless fluid (see, for example, Eq (2.1) of \cite{Blas:2015qsi}). 
In SPT one finds the solution to these equations as a series over the initial field $\delta_0$,
\begin{equation}
\label{eq:expSPT}
\delta_\eta(\textbf{k})=g(\eta)\delta_0(\textbf{k})+g^2(\eta)\int d^3q F_2(\textbf{q},\textbf{k}-\textbf{q})\delta_0(\textbf{q})\delta_0
(\textbf{k}-\textbf{q})+...\,,
\end{equation}
where $g(\eta)$ is the time propagator (growth factor), $F_2$ is the SPT integral kernel. 
Then one can plug the expansion \eqref{eq:expSPT} into Eq.\eqref{eq:Zspt} and find
equal-time correlation functions by taking functional derivatives w.r.t the source $J$. 
It is well known that the expansion \eqref{eq:expSPT} breaks the invariance under the transformations related to the equivalence principle,
which eventually produces spurious IR enhancements in individual SPT diagrams \cite{Scoccimarro:1995if,Creminelli:2013mca}. 
The latter complicates the analysis of physical IR effects (bulk flows) within SPT.

Since only equal-time correlators typically have practical value, 
it seems natural to study the partition function for cosmological fields at particular times.
To do this, one can transform the measure in \eqref{eq:Zspt} $\delta_0\to \delta_\eta(\delta_0)$, which will define a time-dependent 
distribution function of $\delta_\eta$, 
\begin{equation}
\label{eq:Ztspt}
Z[J]= \mathcal{\tilde N}^{-1}\int \mathcal{D}\delta_\eta \; \mathcal{P}[\delta_\eta,\eta]\exp \left\{\int d^3k \;\delta_\eta(\textbf{k})J(-\textbf{k})\right\}\,,
\end{equation}
The equation that governs the time evolution of $\mathcal{P}[\delta_\eta,\eta]$ is a classical Liouville continuity equation, 
\begin{equation}
\label{eq:Liouville}
\frac{\partial}{\partial \eta}\mathcal{P}[\delta_\eta,\eta]+\frac{\partial}{\partial \delta_\eta}\left(\frac{\partial \delta_\eta}{\partial \eta}\mathcal{P}[\delta_\eta,\eta]\right)=0\,.
\end{equation}
One can write $\mathcal{P}[\delta_\eta,\eta]\equiv \exp\left\{-\Gamma[\delta_\eta] \right\}$ and assume, in the spirit of perturbation theory,
\begin{equation}
\Gamma[\delta_\eta]\equiv \sum_{n=1}^{\infty}\frac{1}{n!}\int d^3q_1...d^3q_n  \; \Gamma^{tot}_n(\eta,\textbf{q}_1,...,\textbf{q}_n)\; 
\delta_\eta(\textbf{q}_1)\times ... \times \delta_\eta(\textbf{q}_n)\,.
\end{equation}
Then, using the fluid equations for $\delta_\eta$, one can eliminate ${\partial \delta_\eta}/{\partial \eta}$ 
from Eq.\eqref{eq:Liouville} and arrive at the 
the chain of equations defining the 
time - evolution of the TSPT vertices $\Gamma^{tot}_n$. 
Remarkably, these equations 
can be solved {\it exactly} in the case of Einstein-de Sitter Universe without requiring any truncation of hierarchy.
It is convenient to split $\Gamma^{tot}_n$ into a regular part $\Gamma_n$ and a singular 'counterterm' $C_n$, 
\begin{equation}
\Gamma^{tot}_n=\Gamma_n+C_n\,.
\end{equation}
The terms $C_n$ are related to the singular Jacobian that appears as a result of the measure transformation in Eq.\eqref{eq:Ztspt}.
The vertices $\Gamma_n$ represent physical non-linear n-point couplings and are non-local in space. 
Having imposed Gaussian initial conditions for these vertices
one finds that the time evolution factorizes out, 
\begin{equation}
\Gamma_n(\eta,\textbf{k}_1,...,\textbf{k}_n)=\frac{1}{g^2(\eta)} \delta^{(3)}_{\text{Dirac}}(\textbf{k}_1+...+\textbf{k}_n)\bar \Gamma_n(\textbf{k}_1,...,\textbf{k}_n), \quad \text{where} \quad g(\eta)=D_{+}(\eta)\,,
\end{equation}
and $\bar \Gamma_n$ are time-independent. 
Then one can find cosmological equal-time correlators by
performing a perturbative expansion of \eqref{eq:Ztspt} 
completely analogous to that of quantum field theory.
In this expansion the growth factor $g^2(\eta)$ is treated as a coupling constant,
the initial power spectrum $P(k)~=~\bar \Gamma^{-1}_2(\textbf{k},-\textbf{k})$
plays the role of the free propagator, and higher vertices $\bar \Gamma_n$ ($n>2$)
represent connected parts of the amputated tree-level n-point correlation functions.

Remarkably,  all TSPT vertices are finite in the soft limit, i.e. when some of their arguments are soft, $\lim_{\epsilon \to 0}\bar \Gamma_n(\eta,\textbf{k}_1,...,\textbf{k}_m,\epsilon \textbf{q}_1,...,\epsilon\textbf{q}_{n-m})~=~O(\epsilon^0)$. This property is related to the fact that TSPT operates only equal-time quantities, which are protected by the equivalence principle. Thus, each individual TSPT diagram is IR safe in contrast to SPT.

\begin{figure}
\begin{center}
  \includegraphics[height=4.65cm]{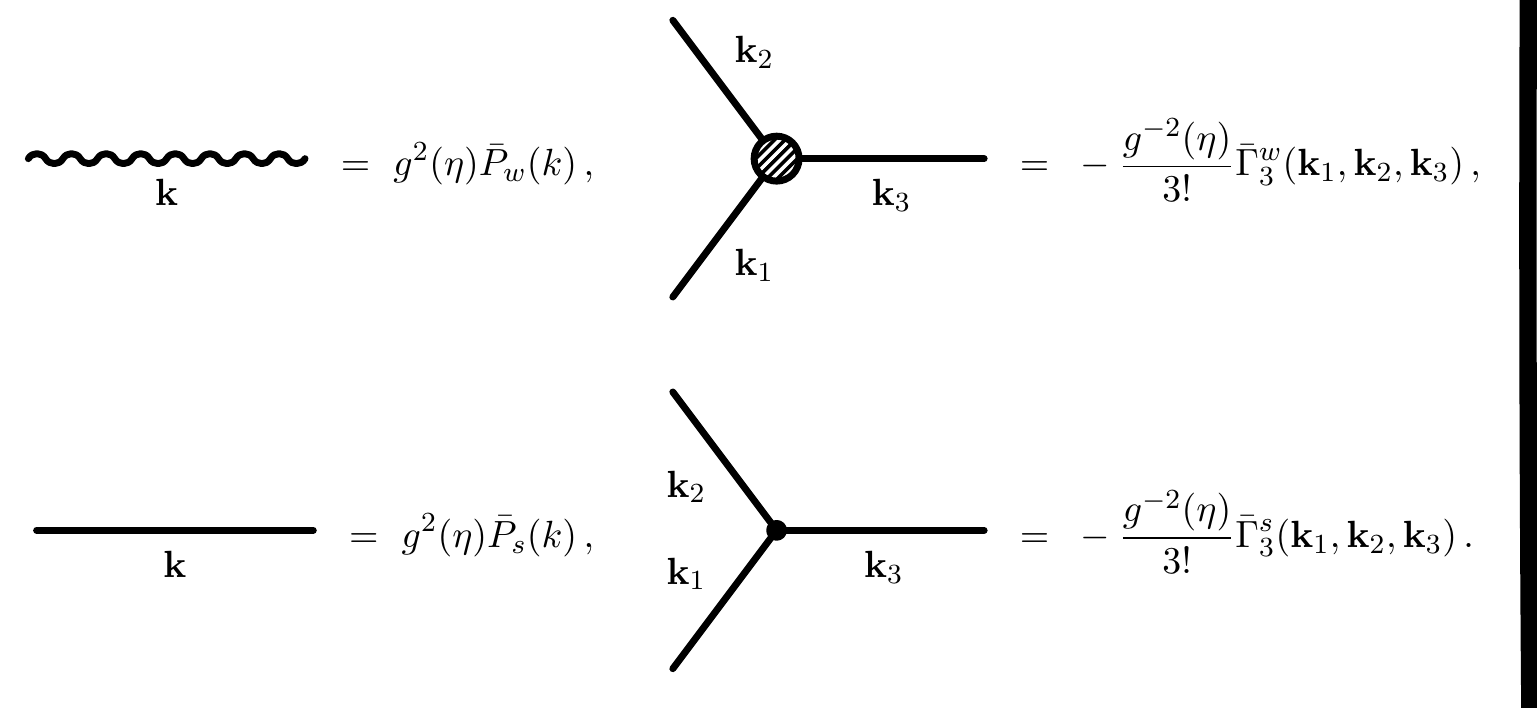}
  \unskip\ \vrule\
    \includegraphics[height=4.65cm]{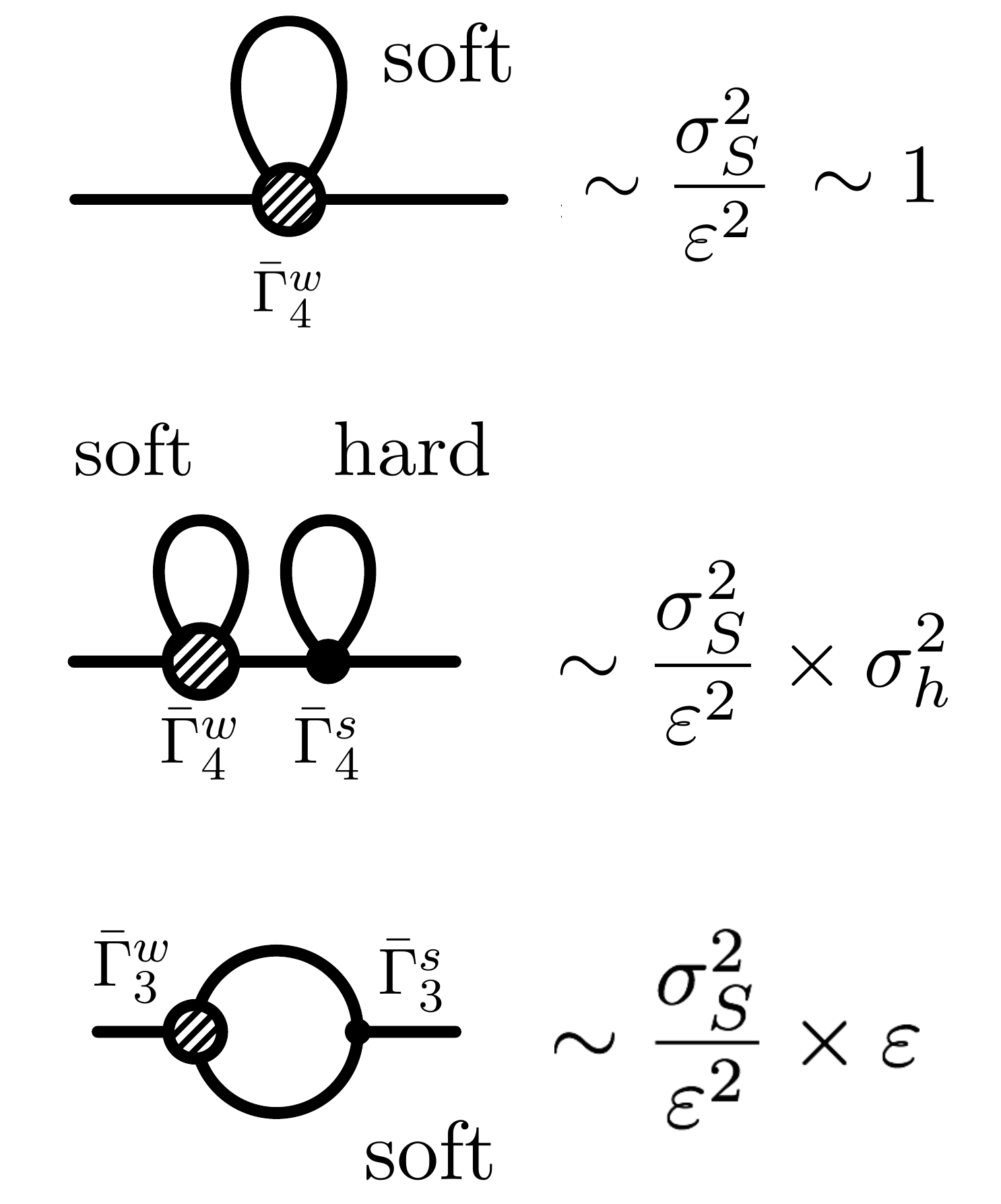}
\end{center}
\caption{\label{fig2} 
{\it Left panel:} Examples of Feynman rules for wiggly and smooth elements. 
{\it Right panel:} Power-counting rules for some TSPT diagrams. 
}
\end{figure}

\section{IR - resummation in TSPT}
\label{sec:ir-res}

The advantageous IR properties of TSPT can be readily used to account for bulk flows.
As we already pointed out, the linear power spectrum in our Universe can be decomposed into the smooth and wiggly parts.
Similar decomposition can also be done for the TSPT vertices, 
\begin{align}
\bar \Gamma_n(\textbf{k}_1,...,\textbf{k}_n)=\bar \Gamma^s_n(\textbf{k}_1,...,\textbf{k}_n)+\bar \Gamma^w_n(\textbf{k}_1,...,\textbf{k}_n)\,.
\end{align}
Thus, the TSPT loop expansion is split into the smooth and wiggly elements separately (see left panel of Fig.~\ref{fig2}), which allows to 
isolate and resumm the relevant for the BAO contributions. One can find that the wiggly vertices are enhanced in the soft limit, e.g.
\begin{equation}
\label{eq:Gnw}
\lim_{q\to 0}\bar \Gamma^w_n(\textbf{k}_1,...,\textbf{k}_m-\sum \textbf{q}_i,\textbf{q}_1,...,\textbf{q}_{n-m})\propto \frac{1}{\varepsilon^{n-m}}\,,\quad 
\text{where} \quad \varepsilon \sim \frac{k_{osc}}{k}\ll 1\,,
\end{equation}
while the smooth vertices do not receive any enhancement in agreement with physical expectations. 
In order to identify and resumm the enhanced contributions at the diagrammatic level one can establish power-counting rules that determine the order of an arbitrary L - loop diagram.
For that one has to set up a 
separation scale $k_S$ which splits the loop integrals into the soft and hard parts. This scale is a priory arbitrary and any residual dependence on it 
should be considered a theoretical error. Then, for each diagram containing a wiggly vertex (other diagrams are not IR-enhanced) one has to:
(1) Assign each loop to be hard ($q>k_S$) or soft ($q<k_S$). Its value then can be associated
with the variance of the density field taken 
over the corresponding domain,
\[
\sigma^2_{S}\equiv \int_{0}^{k_S} d^3q P_s(q)\,,\quad \text{or}\quad \sigma^2_{h}\equiv \int_{k_S}^{\Lambda_{UV}} d^3q P_s(q)\,.
\]
(2)  Count the number of soft legs which are attached to a wiggly vertex. According to Eq.\eqref{eq:Gnw} this number is equal to the level of enhancement due to IR modes.

\begin{figure}
\begin{center}
  \includegraphics[width=0.7\textwidth]{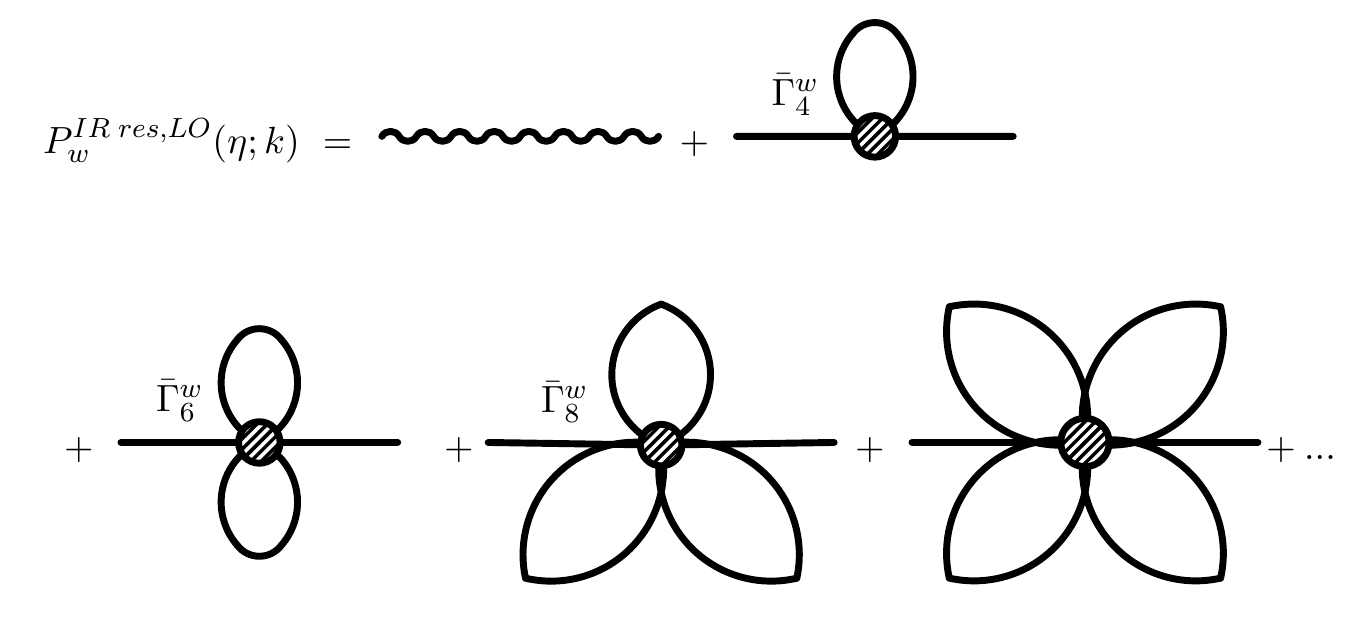}
\end{center}
\caption{\label{fig3} 
Diagrammatic representation of IR resummation at the leading order.
}
\end{figure}

Examples of our power counting rules are shown in the right panel of Fig. \ref{fig2}.
It is easy so see that the leading contributions ($\sim (\sigma^2_S/\varepsilon^2)^L$, L-loop order) 
to the wiggly power spectrum are given by the so-called 'daisy' graphs, 
whose resummation has a simple diagrammatic representation shown in Fig.\ref{fig3}.
One can show that the daisy diagrams exponentiate,
\begin{equation}
\label{eq:LO}
P^{IR\;res,\;LO}_w(\eta;k)=e^{-\Sigma^2k^2}P_w(\eta;k)\,,\quad \text{where}\quad \Sigma^2(\eta,k_S)\equiv \frac{4\pi}{3}\int_0^{k_S}dq P_s(q)\Bigg[1-j_0\left(\frac{q}{k_{osc}}\right)+2j_2\left(\frac{q}{k_{osc}}\right)\Bigg]\,,
\end{equation}
and $j_0(x),j_2(x)$ are spherical Bessel functions.
Remarkably, at leading order the result is the same for exact dynamics and the Zel'dovich approximation.

At next-to-leading order we resumm the corrections involving a hard loop ($\sim$~$(\sigma^2_S/\varepsilon^2)^L\sigma^2_h$),
and the diagrams which contain sub-leading in $\varepsilon$ interaction with the soft modes ($\sim$~$(\sigma^2_S/\varepsilon^2)^L\varepsilon$).  
The result is given by,
\begin{align}
\nonumber
P^{IR\;res,\;NLO}(\eta;k)=&P_s(\eta;k)+e^{-\Sigma^2k^2}P_w(\eta;k)(1+\Sigma^2k^2)\\
\label{eq:NLO}
&+P_{\text{1-loop}}[P_s+e^{-\Sigma^2k^2}P_w]+k^3\Sigma^2_{\text{NLO,soft}} e^{-\Sigma^2k^2}\frac{d}{d k}P_w(\eta;k)\,,
\end{align}
where $\Sigma^2_{\text{NLO,soft}} =\Sigma^2_a+\Sigma^2_b$ has pretty complicated form which can be found in Sec.7 of \cite{Blas:2016sfa}.
Notice that the final NLO IR resummed formula differs quite 
a lot from the simple form of exponential damping.

\section{Comparison with N-body data}

\begin{figure}
\begin{center}
\includegraphics[width=0.49\textwidth]{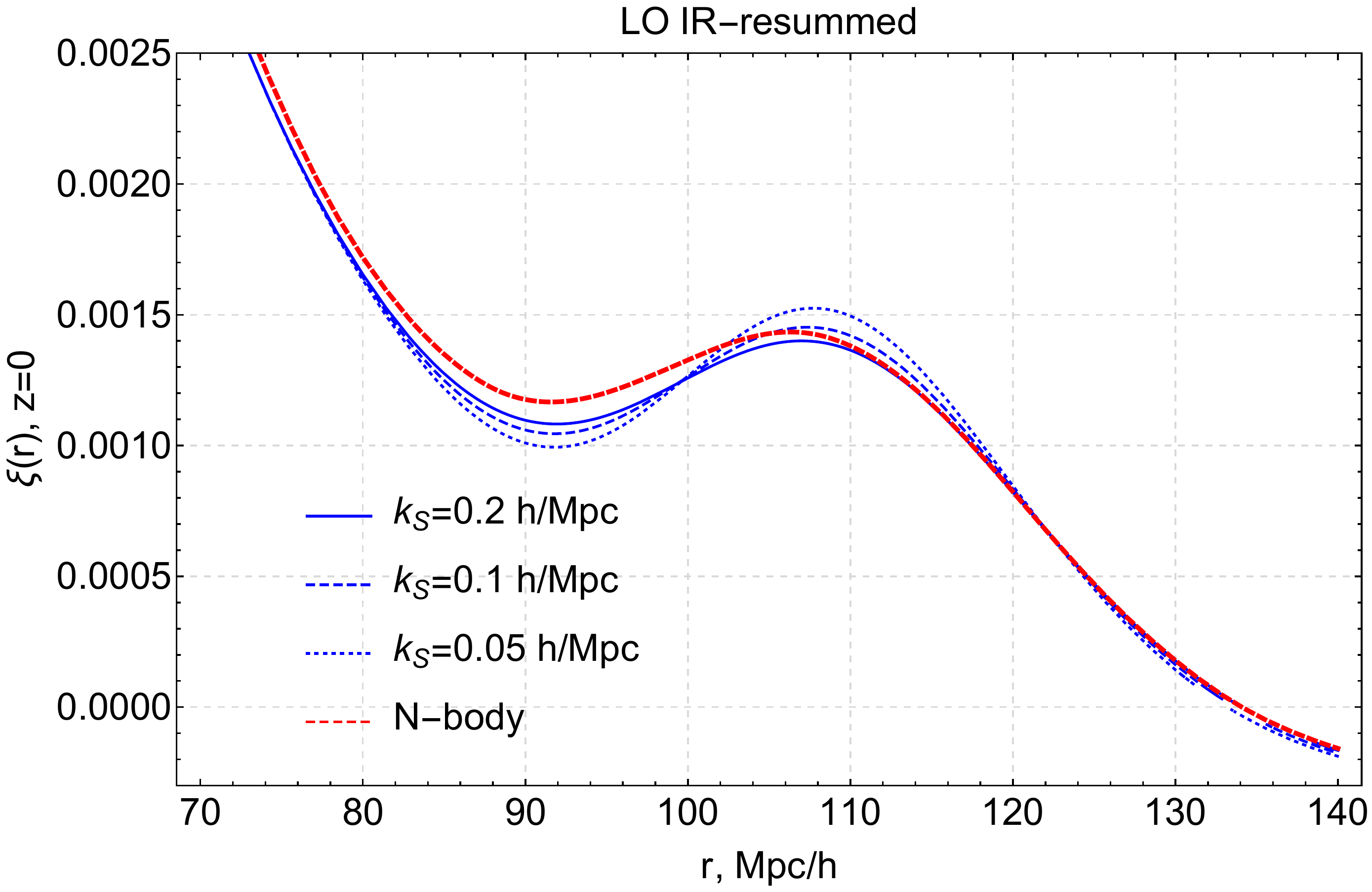}
\includegraphics[width=0.49\textwidth]{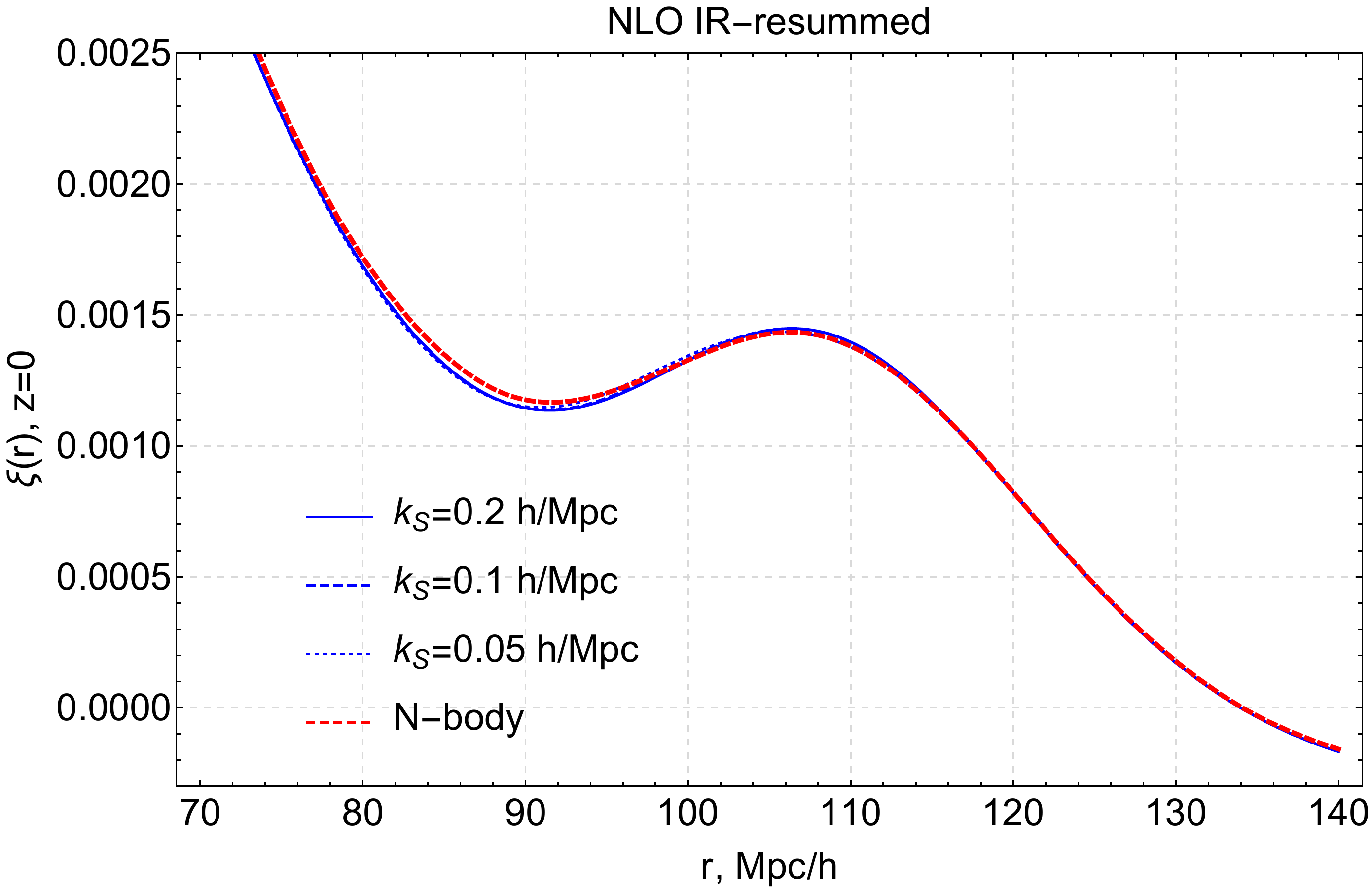}
\end{center}
\caption{\label{fig5} 
IR-resummed matter correlation function at leading (left panel) and next-to-leading (right panel) order. 
N-body data are taken from the Horizon Run 3 simulation \cite{Kim:2011ab}.
}
\end{figure}

In this section we compare our theoretical predictions with the 
N-body simulations Horizon Run 2 and 3 \cite{Kim:2011ab}.
For that we evaluate the matter correlation function in position space, 
\begin{equation}
\xi(r,z)=\frac{4\pi}{r}\int_0^{\infty}dk\;kP(z;k) \sin(kr)\,,
\end{equation}
using the LO and NLO IR - resummed power spectra Eq.\eqref{eq:LO} and Eq.\eqref{eq:NLO} as an input.
The result is shown in Fig.~\ref{fig5}. On the left panel we display the LO IR - resummed result. 
The dependence on the separation scale, which we assume to be a theoretical error, is quite strong in this case. 
In the right panel of Fig.~\ref{fig5} we show the NLO IR - resummed correlation function. 
The agreement with the data is significantly improved as compared to LO. 
Also the dependence on the separation scale is reduced. 
This is an important observation because any dependence on $k_S$ must vanish, in principle, in the exact result.
Thus, the reduction of the theoretical error when going 
to next-to-leading order ensures the consistency of our resummation scheme.
\begin{figure}
\begin{center}
\includegraphics[width=0.59\textwidth]{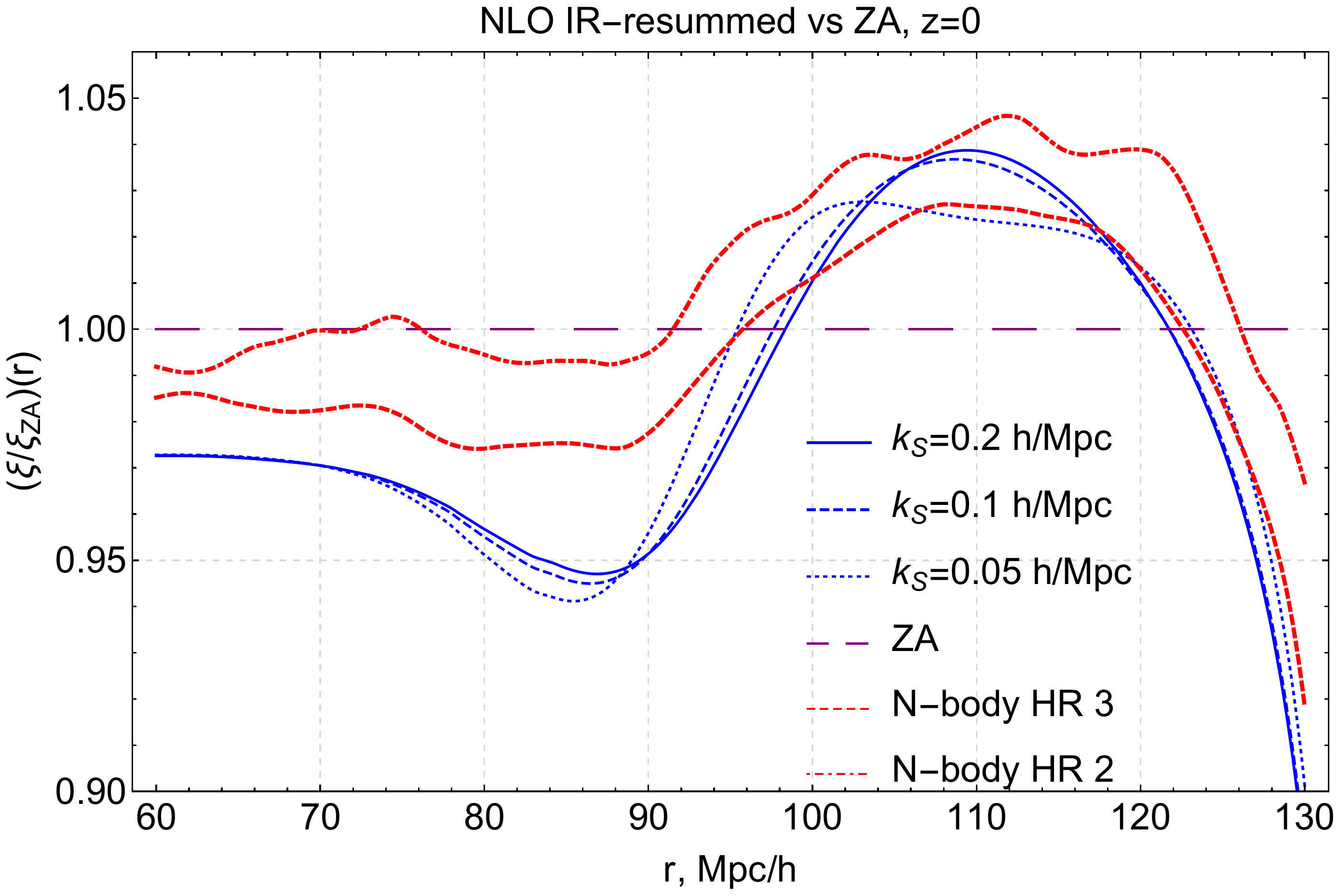}
\end{center}
\caption{\label{fig6} 
IR-resummed matter correlation function at NLO divided by the Zel'dovich approximation. 
Also shown are N-body data from the Horizon Run 2 and 3 simulations \cite{Kim:2011ab}
divided by ZA.}
\end{figure}
It is useful to compare the TSPT NLO predictions with the Zel'dovich approximation (ZA), which is known to be
a quite accurate model for the BAO feature. In Fig.~\ref{fig6} we show the NLO correlation function divided by ZA, along with the 
N-body data Horizon Run~2~and~3. 
The difference between our results and ZA is of order $5\%$ at the BAO peak,
and it seems that the NLO result performs slightly better there.
We notice, however, that the difference is comparable with the 
uncertainties in the N-body data in this region\footnote{Ref.\cite{Kim:2011ab} does not give error bars. 
An estimate for the statistical variance and the resolution of these simulations suggest the error bars of  
order few percent at the BAO peak. 
This level of accuracy is also consistent with the difference between the correlation functions of Horizon Run 2 and Horizon Run 3.},
thus an apparent improvement over ZA seen in the Horizon Run data should be taken with a grain of salt.
Nevertheless, the difference between TSPT NLO and ZA is larger than the forecasted precision of future surveys. 
The TSPT framework can also be used to systematically incorporate the NNLO corrections and 
thus provides a reliable tool for going beyond ZA.

Another non-linear effect on the BAO is the shift of the BAO peak \cite{Sherwin:2012nh,Smith:2007gi}.
This effect is quite small $\sim -0.3\%$ for the $\Lambda$CDM, 
but can be measured with future observations \cite{Weinberg:2012es}.
The shift of the BAO peak may be big in modified gravity models \cite{Bellini:2015oua},
which makes it an interesting observable for constraining the deviations from 
$\Lambda$CDM. 
The shift is induced by terms which are
off-phase with the linear BAO oscillations (see \eqref{eq:SPT1loopIRw}). Those correspond to NLO IR contributions in
our power counting (scale as $\sim (\sigma^2_S/\varepsilon^2)^L\varepsilon$),
and, in principle, can be explicitly extracted from Eq.\eqref{eq:NLO} to yield  
$\delta r_{BAO} / r_{BAO} \approx -(0.3\div  0.5)\%$
which is consistent with values measured in simulations, e.g.\cite{Seo:2009fp}.

\section{Conclusions}
\label{sec:conclus}

In these notes we discussed the non-linear effects in the BAO and sketched the way how one can systematically
take them into account within time-sliced perturbation theory. 
We outlined the physical picture behind the interactions with large scale bulk flows  
and argued the need for IR resummation if one works within the Eulerian framework.
Then we made a short introduction into the TSPT formalism and discussed its key virtues relevant  
for IR resummation: a clear way to separate the perturbative expansion into the smooth and wiggly components, and
manifest IR safety of the TSPT loop integrands. 
We introduced the power counting rules which were used to identify and resumm relevant sets of diagrams at leading and next-to-leading orders.
Finally, we compared our results with N-body data and found good agreement within data errors.

We point out that the TSPT framework can be easily extended to incorporate higher-order corrections due to non-linear clustering, 
as well as new physics, e.g. the effects of neutrino masses or primordial non-gaussianity.

{\bf Acknowledgments} Author is grateful to D. Blas, M. Garny and S. Sibiryakov for their contribution to this work. The work is supported by
the Swiss National Science Foundation.

\end{document}